\documentclass[aps,reprint,twocolumn,amssymb,superscriptaddress,showpacs]{revtex4-1}
\usepackage[]{fontenc}
\usepackage[latin9]{inputenc}
\usepackage{units}
\usepackage{amstext}
\usepackage{amssymb}
\usepackage{graphicx}

\makeatletter
\@ifundefined{textcolor}{}
{%
 \definecolor{BLACK}{gray}{0}
 \definecolor{WHITE}{gray}{1}
 \definecolor{RED}{rgb}{1,0,0}
 \definecolor{GREEN}{rgb}{0,1,0}
 \definecolor{BLUE}{rgb}{0,0,1}
 \definecolor{CYAN}{cmyk}{1,0,0,0}
 \definecolor{MAGENTA}{cmyk}{0,1,0,0}
 \definecolor{YELLOW}{cmyk}{0,0,1,0}
}

\makeatother

\usepackage[english]{babel}
\usepackage{ragged2e}
\begin{document}

\title{Optomechanics with a polarization non-degenerate cavity}

\author{F.M. Buters}
\email{buters@physics.leidenuniv.nl}
\affiliation{Huygens-Kamerlingh Onnes Laboratorium, Universiteit Leiden,
2333 CA Leiden, The Netherlands}
\author{M.J. Weaver}
\affiliation{Department of Physics, University of California, Santa Barbara,
California 93106, USA}
\author{H.J. Eerkens}
\affiliation{Huygens-Kamerlingh Onnes Laboratorium, Universiteit Leiden,
2333 CA Leiden, The Netherlands}
\author{K. Heeck}
\affiliation{Huygens-Kamerlingh Onnes Laboratorium, Universiteit Leiden,
2333 CA Leiden, The Netherlands}
\author{S. de Man}
\affiliation{Huygens-Kamerlingh Onnes Laboratorium, Universiteit Leiden,
2333 CA Leiden, The Netherlands}
\author{D. Bouwmeester}
\affiliation{Huygens-Kamerlingh Onnes Laboratorium, Universiteit Leiden,
2333 CA Leiden, The Netherlands}
\affiliation{Department of Physics, University of California, Santa Barbara,
California 93106, USA}

\date{\today{}}

\begin{abstract}
Experiments in the field of optomechanics do not yet fully exploit the
photon polarization degree of freedom. Here experimental results for
an optomechanical interaction in a polarization non-degenerate system are
presented and schemes are proposed for how to use this interaction to
perform accurate side-band thermometry and to create novel forms of
photon-phonon entanglement. The experimental system utilizes the
compressive force in the mirror attached to a mechanical resonator to
create a micro-mirror with two radii of curvature which leads, when
combined with a second mirror, to a significant polarization splitting of
the cavity modes.
\end{abstract}

\maketitle
Coupling mechanical motion to electromagnetic radiation lies at the heart of cavity optomechanics. Because the coupling is so general, a wide variety of of experimental set-ups exist. For example the scale on which the mechanical motion takes place can range from suspended macroscopic mirrors to cold atoms coupled to an optical cavity \cite{corbitt2007all,borrielli2015low,sankey2010strong,brooks2012non}. Also the source of electromagnetic radiation varies greatly, ranging from the microwave to the optical domain \cite{teufel2011sideband,singh2014optomechanical,chan2011laser,weis2010optomechanically,groblacher2009observation,lu2015squeezed}. 
Each device and set-up has its own advantages. In the optical domain, the availability of the polarization degree of freedom adds an additional knob for controlling and tuning the optomechanical devices. This means that techniques and methods from several landmark  experiments demonstrating photon-photon or photon-matter entanglement can be implemented in existing optomechanical set-ups. However, so far polarization has mostly been used to experimentally separate different optical signals. It has not yet been considered as a degree of freedom in proposals and experiments involving photon-phonon entanglement \cite{marshall2003towards,paternostro2007creating,vitali2007optomechanical,hofer2011quantum,sekatski2014macroscopic,palomaki2013entangling,riedinger2016non}.

This is understandable since the mechanical mode in an optomechanical system is not sensitive to the polarization of the incoming photon. However, the optical mode can be engineered to be polarization sensitive. Birefringence or astigmatism can cause a polarization splitting of the (fundamental) mode of an optical cavity. Although such birefringence has been observed before in optomechanical set-ups, it has been regarded as a parasitic effect \cite{groblacher2009demonstration,jayich2011resolved}. 

In this article we will discuss the possibilities that a polarization non-degenerate cavity has to offer. In some respects using two polarization modes is similar to using two separate laser frequencies. Here we will discuss some of the similarities and differences between existing methods and the method involving multiple polarization modes in the context of two-tone driving, side-band thermometry and experiments close to the mechanical ground state. Finally we will discuss how a polarization non-degenerate cavity can be constructed using a trampoline resonator and show some initial experimental results.  

\begin{figure}
\centering{}\includegraphics[scale=0.36]{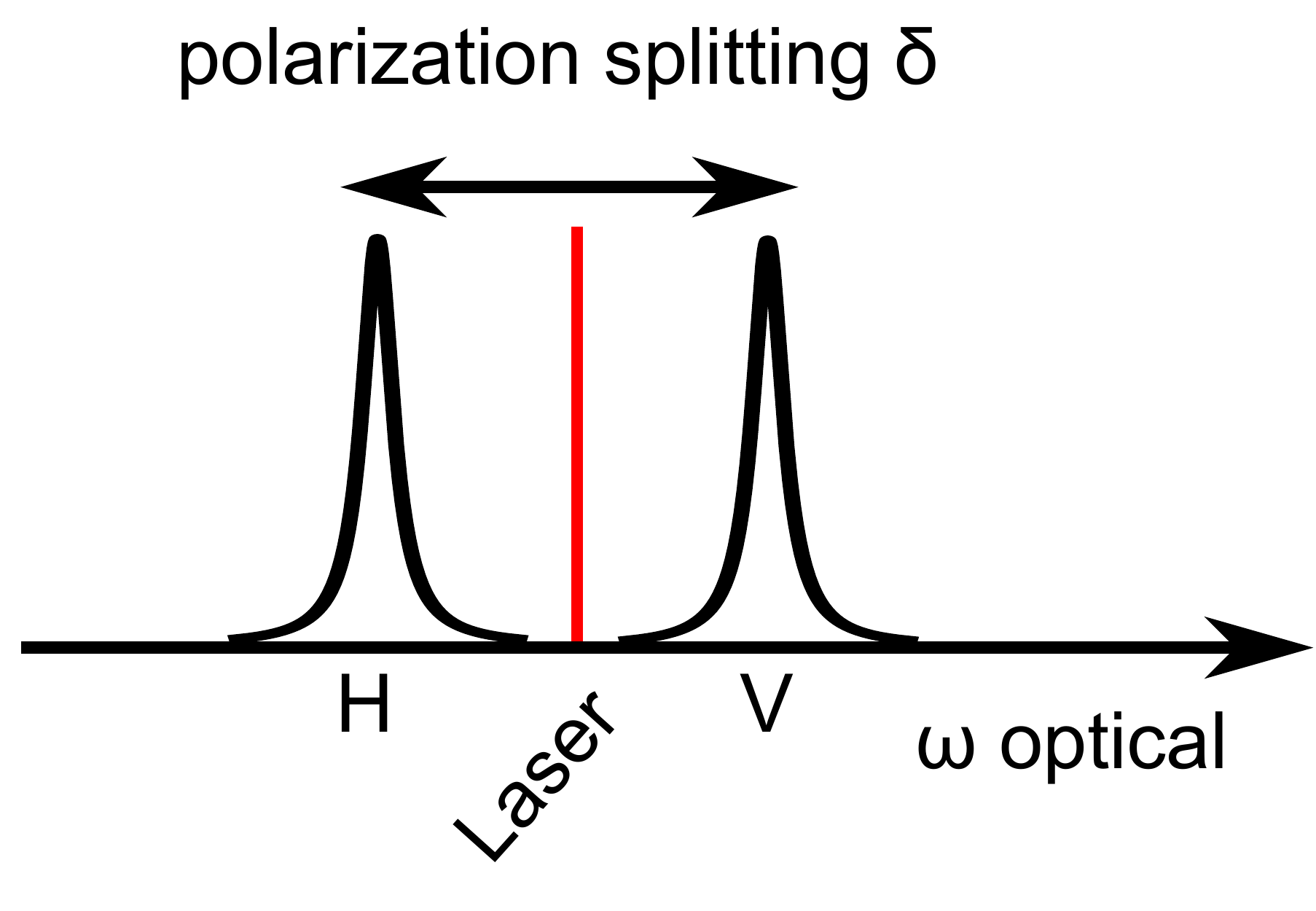}\caption{\label{fig1} Polarization splitting leads to two distinct optical modes when the splitting $\delta$ is larger than the cavity linewidth $\kappa$. H and V denote the different polarization and corresponding cavity mode.}
\end{figure}

Consider the situation depicted in Fig. \ref{fig1} where a laser, 45$^{\circ}$ linearly polarized, is placed in between both cavity modes. We assume the system to be side-band resolved and the polarization splitting to be of the order of twice the mechanical frequency. Projecting the input laser onto the horizontally polarized mode means the laser is blue detuned, while projecting onto the vertically polarized mode results in red detuning. Therefore with a single laser frequency simultaneous driving and cooling can be achieved. Of course the same result is obtained by using two separate laser frequencies. An important difference is that the two polarization modes are orthogonal, while with the two-frequency approach the polarization of the two frequencies can be the same. Careful analysis of the output polarization is therefore required, as we will discuss later. Nonetheless, the single laser frequency solution has also its advantages: when derived from a single laser frequency, any common noise cancels. For experiments relying on a two-tone drive, this can be particularly interesting \cite{clerk2008back,hertzberg2010back,Suh13062014,szorkovszky2011mechanical,wollman2015quantum}.

A second interesting situation occurs when the mechanical resonator is cooled to near its ground state. Stokes scattering, in which the photon down-conversion transfers a phonon to the mechanical resonator, scales with $\langle n \rangle +1$ where  $\langle n \rangle$ is the average phonon number. Anti-Stokes scattering, in which the the photon up-conversion extracts a phonon from the mechanical resonator, scales with $\langle n \rangle $. When the average phonon occupation $\langle n \rangle $ is small, the asymmetry in the Stokes and anti-Stokes scattering can be used to determine the phonon number $\langle n \rangle $. Several groups have already demonstrated this technique to measure accurately the phonon number using either multiple laser frequencies or a heterodyne detection technique \cite{ safavi2012observation, weinstein2014observation, Purdy2015a, underwood2015measurement}. Below we will discuss an alternative method for side-band thermometry when the optomechanical system is polarization sensitive.
 
Consider again the situation where a probe beam, 45$^{\circ}$ linearly polarized, is placed in between both optical modes. For large $\langle n \rangle $, both the horizontal and vertical mode will exit the cavity with equal intensity. However, when the phonon occupation number is lowered, Stokes scattering becomes dominant and the light exiting the cavity will be mainly in the horizontal mode. Therefore the phonon number can be accurately obtained by measuring the ratio of light in the horizontal and vertical mode, since this will scale as $1+1/\langle n \rangle$.

A convenient way to do this is to use a linear polarizer in the path of the transmitted cavity signal, such that the polarization axis is aligned with the input polarization state. This will mix the unperturbed input beam together with the scattered light of both the horizontal and vertical mode. Besides a DC component the photo-detector will register two oscillating components: one at the mechanical frequency $\Omega_{m}$ and one at twice the mechanical frequency. The component at $2\Omega_{m}$, the beat signal from the two side-bands, is directly proportional to the displacement of the mechanical resonator. The component at $\Omega_{m}$ arises from beating the Stokes side-band with the input signal but also from beating the anti-Stokes side-band with the input signal. Since the side-bands have opposing signs, the component at $\Omega_{m}$ is not present when the side-bands are of equal strength, as is the case for large $\langle n \rangle $. However, close to the ground state, the cancellation of the signal at $\Omega_{m}$ no longer takes place. Therefore the signal at $\Omega_{m}$ probes directly the side-band asymmetry.  This method can be an alternative for when the multiple frequency or heterodyne technique proves to be technically challenging. 

Lastly, interesting effects also occur when the mechanical resonator is cooled (close) to its ground state. To create photon-phonon entanglement often a beamsplitter is used \cite{marshall2003towards,paternostro2007creating,vitali2007optomechanical,hofer2011quantum,sekatski2014macroscopic}. For a polarization sensitive cavity this is no longer needed. If we again consider the situation of Fig. \ref{fig1}, but replace the laser with a single photon source, we see that entanglement arises when the incoming photon, 45$^{\circ}$ linearly polarized, is projected onto either basis state. By projecting onto the horizontal basis state, a phonon will be added to the mechanical resonator, while projecting onto the vertical basis state will extract a phonon. When, for example, starting from the $\langle n \rangle = 1$ state, entanglement between the ground state and second excited state is achieved. This is not possible when a beamsplitter is used together with multiple laser frequencies. The addition of the polarization degree of freedom has created a new possibility to manipulate the state of the mechanical resonator. Furthermore, additional tools from the quantum optics toolbox can now be used. The input photon can be replaced with polarization entangled photon pairs, where one photon interacts with the resonator while the state of the other photon is monitored.
We must however remark that for such single photon experiments either single photon strong coupling is required or a post selection method has to be implemented \cite{pepper2012optomechanical}.

\begin{figure}
\centering{}\includegraphics[scale=1.0]{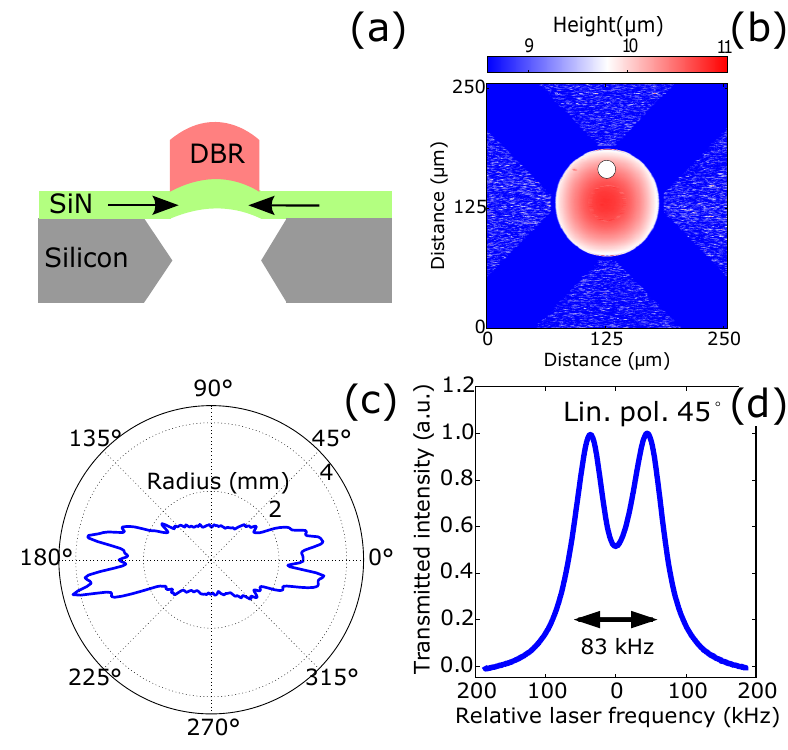}\caption{\label{figbuckle} (a) Compressive stress in the DBR layers causes the mirror to buckle. (b)  Optical profiling with a confocal microscope reveals a concave mirror surface. (c) Local radius of curvature as function of angle obtained for the off center location indicated with the white dot in (a). (d) Demonstration of mode splitting for an off axis aligned cavity by monitoring the transmitted intensity when the laser frequency is varied.}
\end{figure}

All of the above schemes require a polarization non-degenerate optomechanical system. This can be realized when one cavity mirror, either the stationary or the movable mirror, has two radii of curvature. In this work we chose to use the curvature already present in the mirror attached to a trampoline resonator. Finite element analysis using COMSOL shows that the compressive force in the DBR mirror is much larger than the tensile force in the silicon nitride causing the mirror to buckle slightly. This is schematically depicted in Fig. \ref{figbuckle}a. We have confirmed the mirror curvature with an optical profiler. Fig. \ref{figbuckle}b shows a concave mirror surface.
Such small high quality curved mirrors are already interesting on their own to make small micro cavities for cavity QED experiments. For a polarization non-degenerate cavity however, an astigmatic mirror is needed.  
 
Closer inspection of the mirror surface reveals a four-fold symmetry for the curvature in the center of the mirror, as expected from the geometry of the trampoline resonator. Because the DBR mirror is over-sized, 110 $\mu$m diameter, compared to the beam size, typically 12 $\mu$m diameter, a high quality cavity can still be constructed by placing the beam off axis. It is therefore interesting to look at the local curvature away from the middle. For the white dot in \ref{figbuckle}b we determine the local radius of curvature (ROC) by fitting a parabola to a radial line-cut. From the derivative of the parabola the ROC is obtained \cite{Kreyszig1991}. If we repeat this procedure for linecuts at differents angles we obtain Fig. \ref{figbuckle}c. A clear two-fold symmetry is present, with a minimum ROC of about 1 mm and a maximum ROC of about 4 mm. Using these numbers together with the recently published work by Uphoff et al. \cite{uphoff2015frequency}, we expect a polarization splitting of about 60 kHz for the fundamental mode, based on the parameters of our set up.

To demonstrate such a splitting, we constructed a 5 cm long Fabry-Perot cavity operating around 1064 nm and placed in a vibration isolated vacuum chamber. The cavity is aligned to the side of the small curved mirror and the optical linewidth is determined to be 51$\pm$1 kHz by means of a cavity ringdown \cite{eerkens2015optical}. The mechanical resonator is characterized by measuring its mechanical thermal noise spectrum with a laser locked to a cavity resonance using the Pound-Drever-Hall
(PDH) technique \cite{black2001introduction}. With this we measured an intrinsic mechanical linewidth of $\Gamma_{m}=19$ Hz with a mechanical frequency of 222 kHz.

\begin{figure}
\centering{}\includegraphics[scale=0.33]{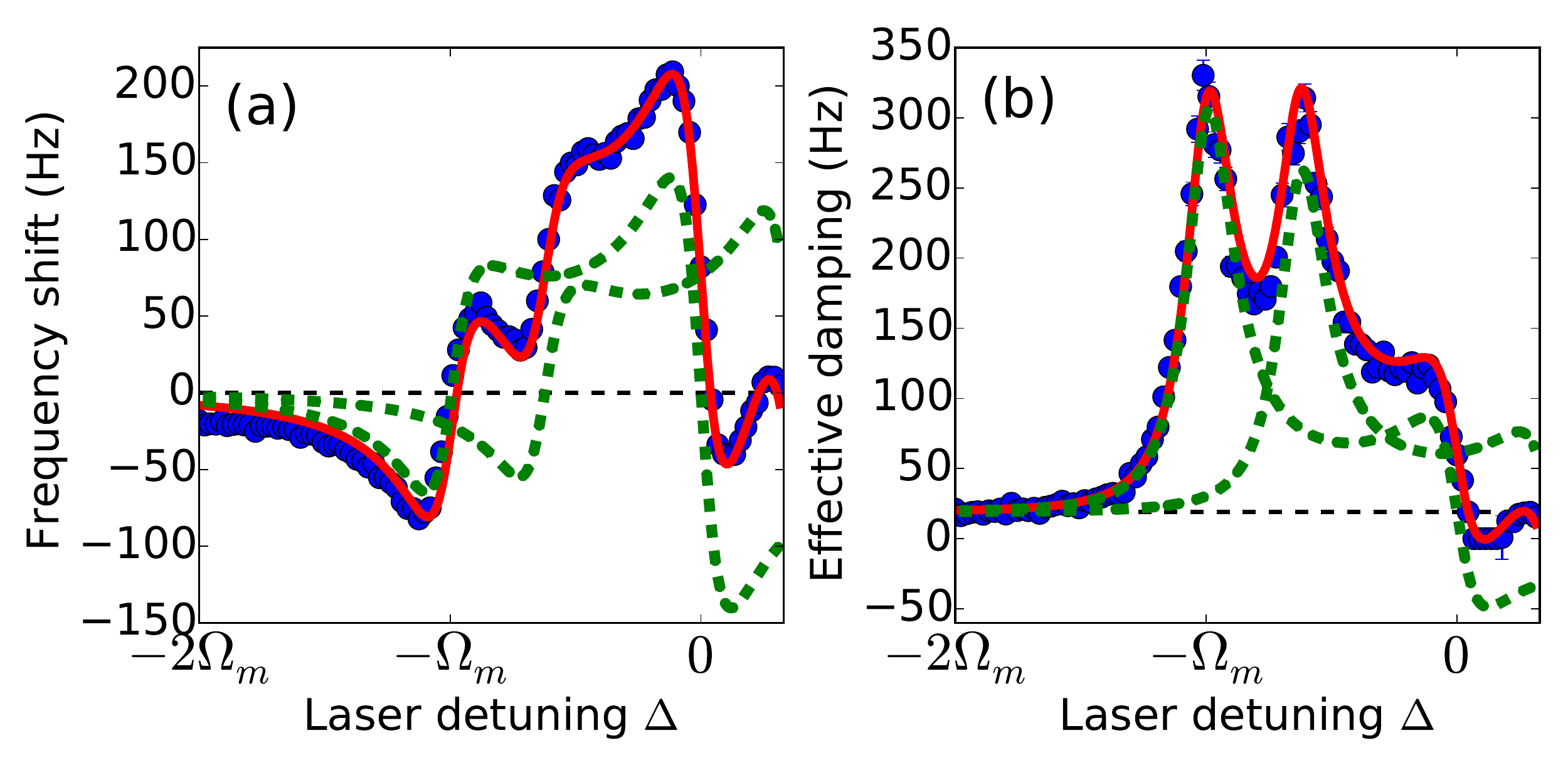}\caption{\label{fig3}(color online) Optomechanical interaction for a polarization non-degenerate cavity. Blue points are extracted from the Lorentzian fit to the mechanical resonance. Red is a simultaneous fit of the data with four free parameters: optical linewidth, optical splitting and input laser power for both modes. Green shows the contribution of the individual modes. (a) Mechanical frequency shift (b) Effective mechanical damping.}
\end{figure}

To see if any polarization splitting is present, we scan the laser across the cavity resonance and adjust the input polarization. A polarization splitting of 83$\pm$1.0 kHz is observed, as shown in Fig. \ref{figbuckle}d. This is of the same order as the expected polarization splitting of 60 kHz. Furthermore, the splitting is large enough to already show some interesting optomechanical effects. For this we use the measurement scheme outlined in ref. \cite{eerkens2015optical}. A probe laser at the cavity resonance is used to monitor the mechanical motion while the detuning of a second pump laser is varied.

\begin{figure}
\centering{}\includegraphics[scale=0.33]{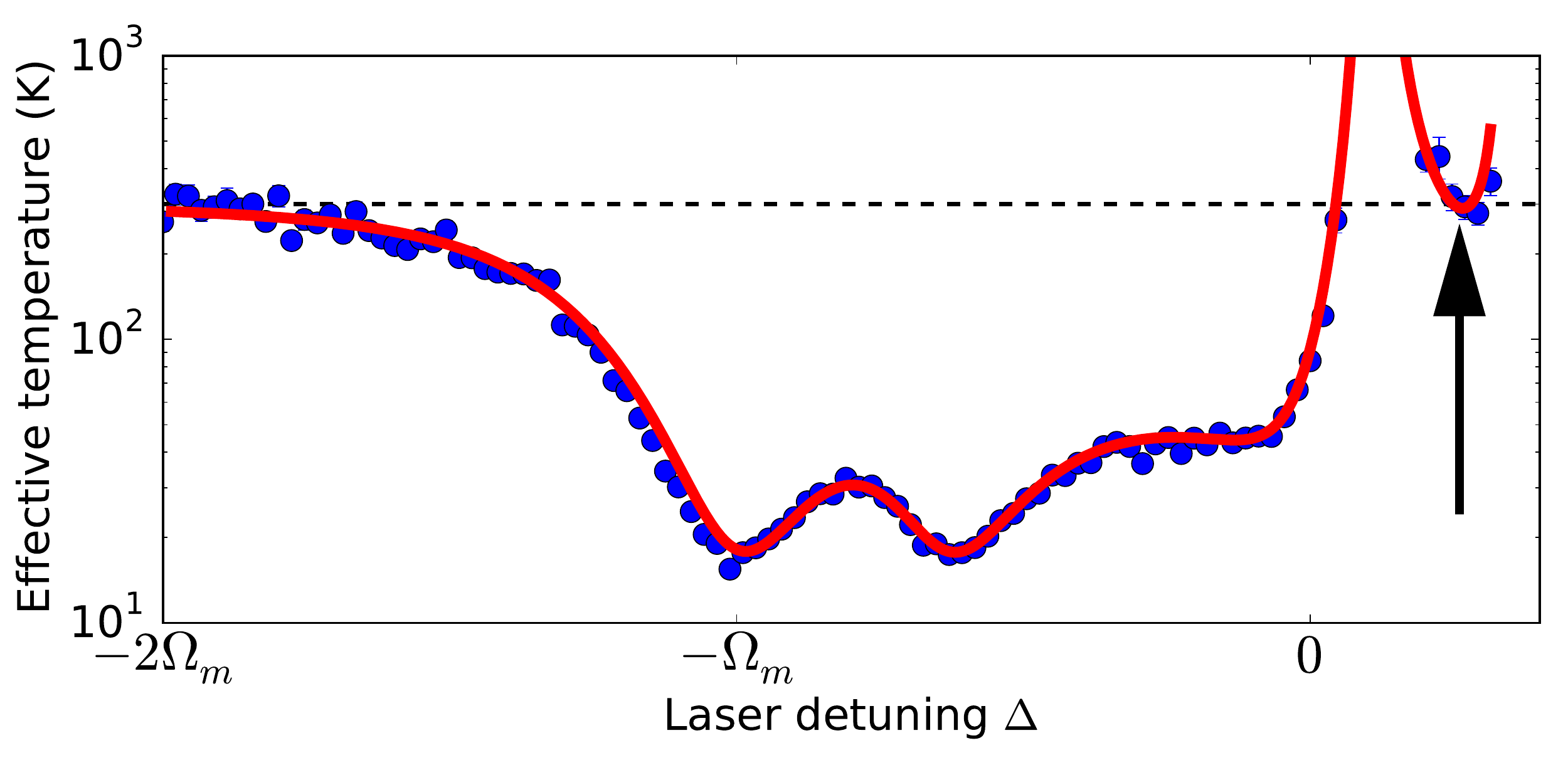}\caption{\label{fig4}(color online) Effective temperature as a function of laser detuning. The arrow indicates the point where the contribution from both modes precisely cancel each other, leading to an effective temperature equal to the bath temperature.}
\end{figure}

For each specific laser detuning we measure the mechanical noise spectrum, fit a Lorentzian and extract the mechanical linewidth and frequency.  The results are shown in Fig. \ref{fig3}. Note that the laser detuning is indicated for one of the two optical modes. The detuning for the other mode is shifted by 83 kHz, the polarization splitting. Since our optomechanical system operates in the linearized regime, we can understand the frequency shift and effective damping by adding  the contributions of both modes:
\begin{eqnarray}
\delta\Omega_{m,total} = \delta\Omega_{m,1} + \delta\Omega_{m,2} \label{eq:eq9}\\
\Gamma_{eff,total} = \Gamma_{opt,1} + \Gamma_{opt,2} + \Gamma_{m} \label{eq:eq10}
\end{eqnarray}
where $\delta\Omega_{m,i}$ and $\Gamma_{opt,i}$ are the optically induced frequency shift and damping (see for example ref. \cite{aspelmeyer2014cavity} for detailed expressions). In green the individual contribution from each mode is shown and in red the result of a fit for the combined effect of both modes. Note that the red curve in Fig. \ref{fig3} and Fig. \ref{fig4} is obtained from a single simultaneous fit to all data with only four free parameters: the optical linewidth, the mode splitting and the input power of both the horizontal and vertical polarization mode.
From Fig. \ref{fig3} we see that the experimental results are nicely described by the addition of the two separate contributions. Furthermore we obtain an optical linewidth $\kappa$ of 52$\pm$0.9 kHz, a mode splitting of 82.4$\pm$1.2 kHz and an input laser power of 2.19$\pm$0.04 $\mu$W and 1.85$\pm$0.04 $\mu$W for both optical modes. These results are in good agreement with the optical characterization. It is also clear that at $\Delta = 41.5$ kHz, precisely in between both optical modes, their contributions cancel. 

This is even more clear when looking at the effective temperature of the mechanical mode, which is obtained from the area of the Lorentzian fit to the mechanical resonance. The theory curve for Fig. \ref{fig4} is given by
\begin{equation}
T_{eff}=\frac{\hbar \Omega_m}{k_B}\frac{\bar{n}_{th}\Gamma_m+\bar{n}_{min}\Gamma_{opt} }{\Gamma_m+\Gamma_{opt}}\label{eq:eq11}
\end{equation}
with $\bar{n}_{min}=\left(\kappa/4\Omega_m\right)^2$ the theoretical minimum phonon number in the side-band resolved regime and $\bar{n}_{th}$ the thermal phonon occupation number. For the optical damping $\Gamma_{opt}$ we use the sum of the contributions from both modes (see Eq. \ref{eq:eq10}).
From the resulting graph of Fig. \ref{fig4} we see again that the experimental results follow the theory nicely. Furthermore, at laser detuning $\Delta = 41.5$ kHz indicated by the arrow, the effective mode temperature is just the environmental temperature, showing once more that the contribution from both polarization modes cancels out.

Although the experimental results shown here are not yet in the target regime whereby the polarization splitting is two times the mechanical frequency, it does show how the mechanical resonator can be simultaneously driven and cooled using a single driving tone. The remaining step is to tune the mechanical frequency and the polarization splitting.

We are already able to produce optomechanical cavities with an optical linewidth smaller than 17 kHz, a mechanical frequency of 250 kHz and an mechanical quality factor approaching 500k \cite{weaver2015nested}. If we take this as a starting point, we need to create a cavity with a mode splitting of 500 kHz. Assuming with some small modifications to the trampoline resonator design the mode splitting can be pushed from 83 kHz to about 100 kHz, the cavity length needs to be reduced by a factor of five, since the mode splitting scales inversely with cavity length \cite{uphoff2015frequency}. This will also increase the optomechanical coupling strength g$_{0}$ to about  $2 \pi \times$ 8 rad/s. A downside to this method is that the cavity linewidth increases by a factor of five, but an optical linewidth of 85 kHz is still sufficient to be side-band resolved. More importantly, to achieve groundstate cooling, the multi-photon cooperativity should be much larger than the thermal occupation number. In this case a base temperature of 1 K together with a laser power of 50 $\mu$W is needed, which is experimentally feasible.

In conclusion, we have proposed how a polarization non-degenerate optomechanical system can be a valuable addition to the existing optomechanical toolbox. It allows for alternative methods to realize a two tone cavity drive as well as side-band thermometry. Furthermore at low phonon occupation numbers new possibilities arise for photon - phonon entanglement. Finally we have demonstrated how such a system can be fabricated and showed that the last remaining step is to decrease the length to achieve the desired mode splitting. 

\section*{Funding Information}
This work is part of the research program of the Foundation for Fundamental Research (FOM) and of the NWO VICI research program, which are both part of the Netherlands Organisation for Scientific Research (NWO).
This work is also supported by the National Science Foundation Grant No. PHY-1212483.

\section*{Acknowledgments}

The authors acknowledge the useful discussions with W. Loeffler. The authors would also like to thank H. van der Meer for technical assistance and support.

\end{document}